# Group-velocity-locked vector soliton molecules in a birefringence-enhanced fiber laser

Yiyang Luo[1, 2], Jianwei Cheng[2], Bowen Liu[2], Qizhen Sun[2], Lei Li[1], Songnian Fu[2], Dingyuan Tang[1], Luming Zhao[1], Deming Liu[2]

[1]Jiangsu Key Laboratory of Advanced Laser Materials and Devices, Jiangsu Collaborative Innovation Center of Advanced Laser Technology and Emerging Industry, School of Physics and Electronic Engineering, Jiangsu Normal University, Xuzhou, 221116 Jiangsu, P. R. China;

[2]School of Optical and Electronic Information, National Engineering Laboratory for Next Generation Internet Access System, Huazhong University of Science and Technology, Wuhan 430074, Hubei, P. R. China;

*Corresponding authors: zhaoluming@jsnu.edu.cn.

## Abstract

Physics phenomena of multi-soliton complexes have enriched the life of dissipative solitons in fiber lasers. By developing a birefringence-enhanced fiber laser, we report the first experimental observation of group-velocity-locked vector soliton (GVLVS) molecules. The birefringence-enhanced fiber laser facilitates the generation of GVLVSs, where the two orthogonally polarized components are coupled together to form a multi-soliton complex. Moreover, the interaction of repulsive and attractive forces between multiple pulses binds the particle-like GVLVSs together in time domain to further form compound multi-soliton complexes, namely GVLVS molecules. By adopting the polarization-resolved measurement, we show that the two orthogonally polarized components of the GVLVS molecules are both soliton molecules supported by the strongly modulated spectral fringes and the double-humped intensity profiles. Additionally, GVLVS molecules with various soliton separations are also observed by adjusting the pump power and the polarization controller.

## Introduction

Passively mode-locked fiber lasers have attracted intensive research for the applications of ultrashort pulsed light source with unprecedented pulse energy or repetition rate, as well as the exploration of dynamics of optical dissipative solitons (DSs)[1-3]. Different from the conservative solitons purely originating from the simple balance between dispersion and nonlinearity, continuous flow of energy gets involved in dissipative systems and the composite balance among cavity dispersion, fiber nonlinearity, laser gain and loss of the fiber lasers, plays a significant role in the formation of DSs. Due to the peak power clamping effect, fundamental solitons tend to be unstable after amplification[4], and more complex waveforms such as pulse bunching[5], harmonic mode-locking[6], soliton molecules[7, 8] and soliton rain[9, 10] can appear with increasing pump power. In particular, soliton molecules (i.e. bound states of solitons), featuring multi-soliton structures, have been a topic of intensive research in fiber lasers. In analogy with the physics of matter, soliton molecules as a multi-soliton complex essentially stem from the interaction of repulsive and attractive forces between particle-like solitons; the formation is a dissipative process that requires the supply of binding energy[1]. Meanwhile, quite different from the well-known high-order solitons, the soliton molecules will not decay under infinitesimal perturbation[2], which ensures their promising applications in optical communications for coding and transmission of information in high-level modulation formats, and increasing capacity of communication channels beyond binary coding limits[11-13]. From the first theoretical study of soliton molecules by Malomed[14] and Akhmediev et al.[15], numerous experimental observations have been reported in nonlinear polarization rotation[16], molybdenum disulfide[17], carbon nanotube[18], and figure-of-eight[19] based fiber lasers. In particular, tightly bound states are the most conventional type, which are interpreted in terms of fixed and discrete soliton separations. Additionally, various types of soliton molecules, such as vibrating soliton molecules, oscillating soliton molecules[1, 20], soliton molecules with flipping and independently evolving phase[21, 22,] have been also theoretically predicted and experimentally observed.

If no polarization discrimination component is used in fiber lasers, the two orthogonally polarized components could be coupled together to form another type of multi-soliton complexes, namely vector solitons in fiber lasers, which not only possess much more plentiful behaviors and richer dynamics than their scalar counterparts, but also pave a promising way for numerous applications from nano-optics to high-capacity fiber optic communications[23-25]. S. T. Cundiff et al. first demonstrated that both the group-velocity-locked vector solitons (GVLVSs) with rotating polarization state and the phase-locked vector solitons (PLVSs) with fixed polarization state could be formed in fiber lasers[26, 27]. L. M. Zhao et al. further revealed that with moderate linear cavity birefringence, another type of GVLVSs with a unique feature of two sets of Kelly sidebands could be formed in fiber lasers[28]. In particular, the two orthogonally polarized components of such vector solitons shift their central frequency in opposite directions to compensate the birefringence-induced group-velocity difference. Thereupon, these two components can trap each other and travel as a unit in time domain. Especially, the two set of Kelly sidebands are interpreted as the indicator of the central frequency shift; and the stronger birefringence is the laser cavity, the larger wavelength shift of the spectrum. In addition, vector solitons with locked and precessing states of polarization[29, 30], vector solitons in dispersion-managed regime[31] and high-order PLVSs[32] have been reported in fiber lasers. More recently, the mergence of soliton molecules and vectorial nature of light spreads a new concept of vector soliton molecules that inspires us to reveal more interesting behaviors and underlying dynamics of these compound multi-soliton complexes formed at both time domain and polarization directions[33-35]. It is found that vector solitons as a non-dispersive unit can also attract or repel each other to form the vector soliton molecules, which provide a possibility for twofold increasing the optical communication capacity.

In this paper, we report an experimental manifestation of GVLVS molecules in a birefringence-enhanced fiber laser. With enhanced net cavity birefringence introduced by a segment of polarization-maintaining fiber (PMF), GVLVSs interpreted as a multi-soliton complex are formed. The two orthogonally polarized components of the vector solitons possess obvious central wavelength shift and travel as a non-dispersive unit in the laser cavity at the same group velocity. Furthermore, with appropriate settings, GVLVS molecules as compound multi-soliton complexes are obtained. Polarization-resolved measurement is adopted to explore the vectorial nature of the GVLVS molecules. In addition, GVLVS molecules with various soliton separations are experimentally observed. The proposed fiber laser serves as an ideal testbed for both exploring the dynamics of DSs and extending their potential applications.

## Results

**Experimental setup.** The experimental setup is schematically depicted in Fig. 1. The mode-locking of the ring cavity fiber laser is initialized by a fiber-pigtailed semiconductor saturable absorber mirror (SESAM, BATOP) with saturable absorption of 8%, modulation depth of 5.5% and recovery time of 2 ps, as shown in the inset of Fig. 1. A three-port polarization-independent circulator is used to incorporate the SESAM into the cavity and simultaneously force the unidirectional operation of the fiber laser. A 1.5 m erbium-doped fiber (EDF, OFS EDF80) with group-velocity dispersion (GVD) of about −48 (ps/nm)/km is used as the gain medium, which is pumped by a 980 nm laser diode (LD) via a 980/1550 nm wavelength division multiplexer (WDM). Backward pump scheme is employed to possibly avoid the overdriving of the SESAM by the residual pump power. The net cavity birefringence is enhanced by a 0.15 m PMF, thus leading to an obvious central wavelength shift. A fiber-based polarization controller (PC) is applied between the PMF and the ingoing branch of the circulator to finely adjust the intra-cavity polarization state. A 10:90 fiber output coupler (OC) is utilized to output the laser emission. Pigtails of all the components are standard single-mode fiber (SMF) with GVD of +17 (ps/nm)/km, and the total cavity length is about 21.5 m. In addition, to get insight into the two orthogonally polarized components, a fiber-pigtailed polarization beam splitter (PBS) is connected to the 10% output port of the OC. The incoming branch of the PBS is made of SMF, while its two outgoing branches are made of PMF. To avoid influence of the fiber birefringence introduced by the fiber segments between the OC and the measurement devices, a fiber-based PC is employed to compensate the extra birefringence.

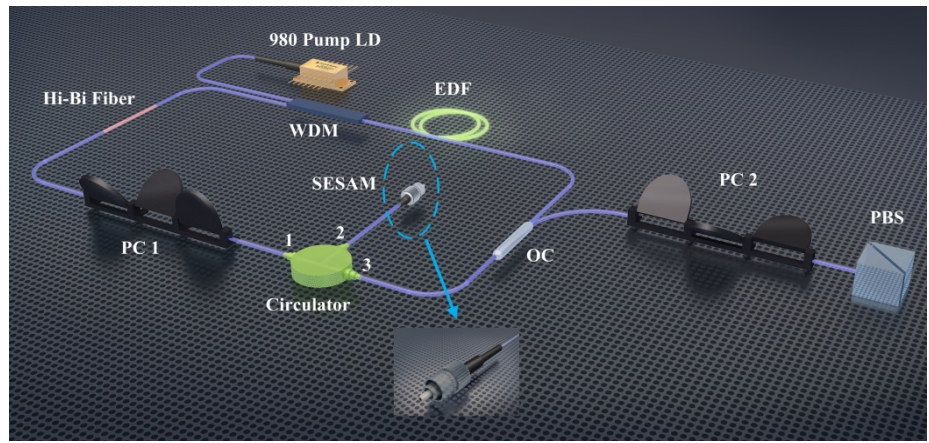

**Figure 1.** Schematic illustration of the experimental setup: passively mode-locked fiber laser and polarization-resolved measurement. LD, laser diode; WDM, wavelength division multiplexer; EDF, erbium-doped fiber; SESAM, semiconductor saturable absorber mirror; PC, polarization controller; OC, output coupler; PMF, polarization-maintaining fiber; PBS, polarization beam splitter.

**Group-velocity-locked vector solitons.** With appropriate settings, self-started mode-locking of the fiber laser is achieved by simply increasing the pump power above the mode-locking threshold, and the mode-locked pulses are automatically evolved into vector solitons as no polarization-sensitive components are adopted in the laser cavity. The enhanced net birefringence of the laser cavity leads to the formation of GVLVSs as illustrated in Fig. 2. Two orthogonally polarized components with obvious central wavelength difference are coupled together to form this multi-soliton complex. The optical spectrum of the GVLVS is shown in Fig. 2(a), where the distinct indicator, namely two sets of Kelly sidebands is clearly observed and the 3-dB spectral bandwidth is about 4.1 nm. The autocorrelation trace depicted in Fig. 2(b) declares a pulse width of 0.69 ps with the assumption of a $sech^2$ pulse shape. The time-bandwidth product is about 0.354, indicating that the output pulse is nearly transformed-limited. Figure 2(c) presents the radio frequency (RF) spectrum of the vector solitons. The fundamental repetition rate is fixed at 9.6 MHz, exactly agreeing with the pulse interval of 105 ns as depicted in Fig. 2(d).

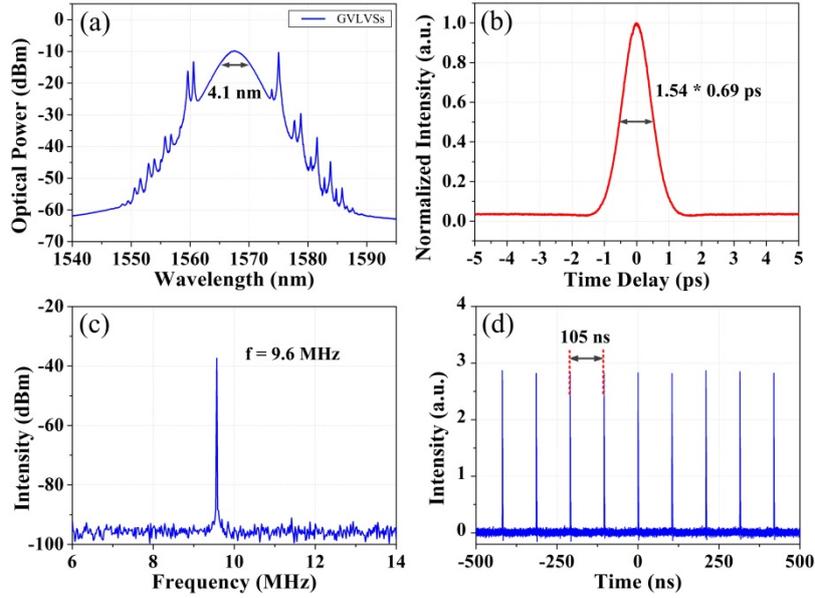

**Figure 2.** Group-velocity-locked vector solitons: (a) Optical spectrum, (b) autocorrelation trace, (c) RF spectrum and (d) fundamental mode-locked oscilloscope trace of the GVLVSs.

From fundamental operation regime, through increasing the pump power, the fiber laser tends to operate at multi-pulse states due to the soliton energy quantization effect. In particular, the dual- and triple-pulse states are respectively shown in Fig. 3(a) and 3(b), corresponding to two pulses and three pulses coexisting in one cavity roundtrip time of 105 ns. The small pulse intensity fluctuation is caused by the detection system. The existence of these independent particle-like solitons provides the possibility for the formation of soliton molecules.

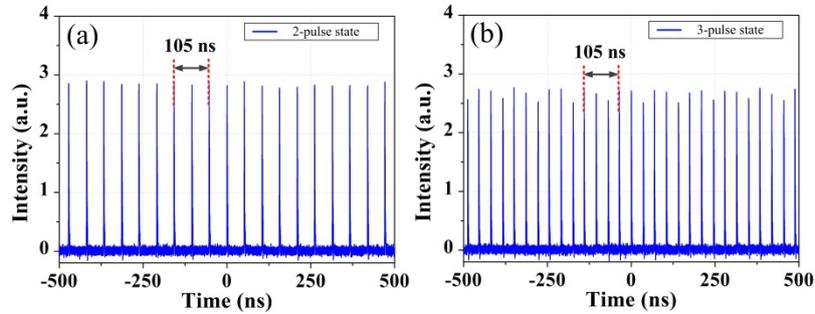

**Figure 3.** Oscilloscope traces of multi-pulse states: (a) Dual-pulse and (b) Triple-pulse state.

**Group-velocity-locked vector soliton molecules.** By carefully manipulating the intra-cavity polarization state, these separate particle-like solitons can be bound together to form soliton molecules due to internal interaction, which are interpreted as another type of multi-soliton complexes formed in time domain. As shown in Fig. 4(a), the distinct indicator of GVLVSs, namely two sets of Kelly sidebands, is also observed from the optical spectrum of the formed soliton molecules. It implies that these soliton molecules stem from the interaction of two GVLVSs, and still have the vector features. The unique feature of the soliton molecules is expressed by the strongly modulated spectral fringes. And the 3.5-nm modulation period manifests that the two GVLVSs are closely spaced. Moreover, the soliton molecules can be further verified by the double-humped intensity profile, as shown in Fig. 4(b). The pulse width of the individual soliton is about 0.81 ps if a $sech^2$ pulse shape is assumed; the soliton separation is estimated to be 2.3 ps, exactly matching the modulation period of the spectral fringes. The soliton separation is approximately 2.8 times of the pulse width, and the height ratio of the three peaks of the autocorrelation trace is around 1:2:1. Thus it can be seen that two GVLVSs of identical intensity are tightly bound together to form the GVLVS molecule. Figure 4(c) presents the corresponding RF spectrum. The fundamental repetition rate maintains at 9.6 MHz. Apart from the fundamental mode-locking state of GVLVS molecules, these compound multi-soliton complexes as a unit can also rearrange themselves in a high-pump regime. In particular, triple-pulse state is confirmed by the oscilloscope trace as shown in Fig. 4(d), where three GVLVS molecules, essentially six GVLVSs coexist in the fiber laser.

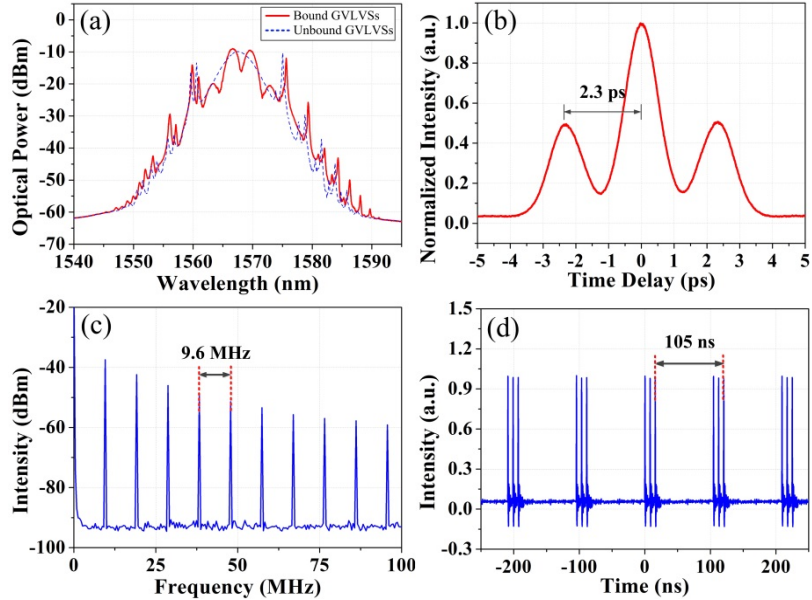

**Figure 4.** Group-velocity-locked vector soliton molecules: (a) Optical spectra of the GVLVSs and GVLVS molecules (dashed blue line and solid red line, respectively); (b) autocorrelation trace and (c) RF spectrum of the GVLVS molecules; (d) oscilloscope trace of the triple-pulse state.

Furthermore, polarization-resolved measurement[28] is adopted to get insight of the vectorial nature of these GVLVS molecules. Figure 5(a) shows the polarization-resolved spectra. 'Total' refers to the direct output of the fiber laser; 'Horizontal axis' and 'Vertical axis' refer to the polarization-resolved output after the PBS, respectively. The polarization-resolved spectra are all strongly modulated with the same period of 3.5 nm. Each of the components along the two orthogonally polarized axes respectively corresponds to an individual set of Kelly sidebands, which is in accordance with the GVLVSs. The slight residual of the sidebands from the other polarization direction results from the low polarization extinction ratio of the PBS. Figure 5(b) depicts the autocorrelation traces of the polarization-resolved components of the GVLVS molecule. The double-humped intensity profiles suggest that the two orthogonally polarized components could be both considered as soliton molecules. They possess almost the same pulse width of 1.08 ps with the assumption of a $sech^2$ pulse shape, and soliton separation of 2.3 ps. Compared to the pulse width of the individual soliton presented in Fig. 4(b), the pulse width of the corresponding polarization-resolved component is broadened from 0.81 ps to 1.08 ps. It is most likely ascribed to the extra dispersion introduced by the pigtails of the PBS. Figure 5(c) presents the polarization-resolved oscilloscope traces of fundamental GVLVS molecules. The pulse intensity of each orthogonally polarized component is identical.

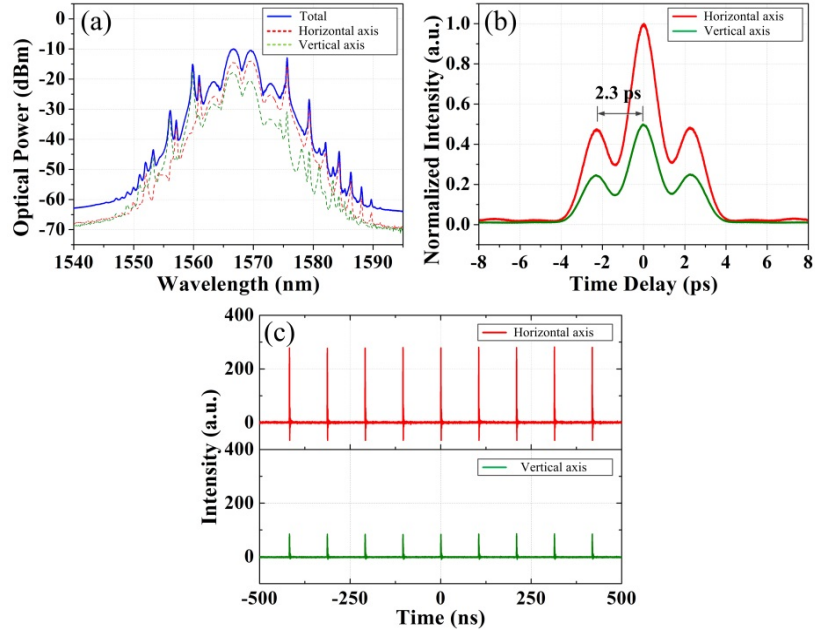

**Figure 5.** Polarization-resolved measurement of GVLVS molecules: (a) Polarization-resolved spectra of the GVLVS molecules; (b) autocorrelation traces and (c) oscilloscope traces of the two orthogonally polarized components of the GVLVS molecules.

**Tightly/loosely bound states.** The adjustment of the pump power and the intra-cavity PC leads to the change of the GVLVSs, which further plays a role in the soliton interaction of repulsive and attractive forces. Hereby, GVLVS molecules with various soliton separations are also observed in this fiber laser. Particularly, GVLVS molecules shown in Fig. 6(a) and 6(b) possess a spectral modulation period of 1.8 nm and a corresponding soliton separation of 4.4 ps, which is around 4.2 times of the pulse width of 1.05 ps if a sech$^2$ pulse shape is assumed. Thus it can be seen that the two GVLVSs are tightly bound together[36]. Additionally, Figures 6(c) and 6(d) exhibit an example of loosely bound state. The soliton separation of 19.6 ps exactly agrees with the spectral modulation period of 0.4 nm, which is almost 19 times of the pulse width of 1.05 ps. Remarkably the autocorrelation traces as shown in Fig. 6(b) and 6(d) display a reduction of the amplitude of the side peak, which are different from that depicted in Fig. 4(b). This phenomenon could be ascribed to the nearby GVLVSs accompanying with the GVLVS molecules in the fiber laser. Our experimental observations signify that the soliton separation could be flexibly controlled by finely adjusting the pump power and the PC.

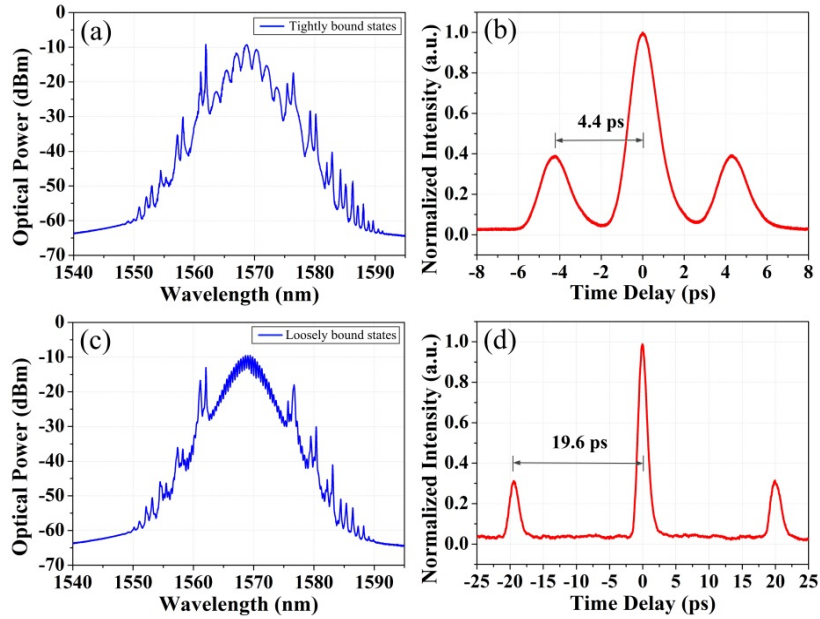

**Figure 6.** Tightly/loosely bound states: (a) Optical spectrum and (b) autocorrelation trace of the tightly bound state; (c) optical spectrum and (d) autocorrelation trace of the loosely bound state.

## Discussions

We develop a birefringence-enhanced fiber laser to explore the new concept of GVLVS molecules. As no explicit polarization sensitive components are adopted and fibers are weakly birefringent in the laser cavity, the formed solitons are essentially vector solitons. The PMF is purposefully utilized in the cavity thus local strong birefringence is introduced. Therefore, the orthogonally polarized solitons experience fast polarization evolution. In particular, birefringence induced polarization dispersion enforces the pulse splitting during propagation in the PMF segment; while solitons along the fast axis and the slow axis shift their central frequencies in opposite directions to slow down and speed up, respectively. In effect, these two orthogonally polarized components create an attractive potential well, thus trapping each other through cross-phase modulation and propagating as a non-dispersive soliton complex at the same group velocity. The formed GVLVSs are validated by the distinct spectrum with two sets of Kelly sidebands.

Next with careful polarization-state manipulation, these separate particle-like GVLVSs can be bound together at time domain to form the compound multi-soliton complexes, namely GVLVS molecules. The GVLVS molecules behave themselves with dual properties. On one hand, the unique feature of soliton molecules is characterized by the strongly modulated spectral fringes, and further verified by the double-humped intensity profile. On the other hand, the GVLVS molecules display distinct spectrum of two set of Kelly sidebands, and their two orthogonally polarized components are both soliton molecules with the identical soliton separation. Besides, by finely adjusting the pump power and intra-cavity polarization state, GVLVS molecules with various soliton separations could be experimentally observed. Overall, the developed birefringence-enhanced fiber laser approaches to the operation regime of GVLVS molecules. The investigation on the interesting behaviors and underlying dynamics of these compound multi-soliton complexes provides the opportunities for both enriching the life of DSs and extending their potential applications.

## Methods

An optical spectrum analyzer (OSA, Yokogawa AQ6370C-20) with a resolution of 0.02 nm and a 1

GHz real-time oscilloscope (OSC, Agilent DSO9104A) together with a 2 GHz photodetector (PD, Thorlabs DET01CFC) are respectively employed to monitor the optical spectra and the pulse train of DSs. RF spectrum is analyzed by a 3 GHz electrical spectrum analyzer (ESA, Agilent N9320B). Finally, the pulse profile is measured by a commercial autocorrelator (Femtochrome FR-103HS).

## Acknowledgement

This work is supported by the sub-Project of the Major Program of the National Natural Science Foundation of China (No.61290315), the National Key Scientific Instrument and Equipment Development Project of China (No. 2013YQ16048707), and the National Natural Science Foundation of China (No. 11674133, No. 61275004, No. 61575089).